\documentclass[aps,prl,amsmath,twocolumn]{revtex4}
\usepackage{graphics, setspace,epsfig,color,subfigure}
\usepackage[vcentermath]{}

\newcommand{\be}{\begin{equation}}
\newcommand{\ee}{\end{equation}}

\begin{document}

\title[title]{Is there a ``most perfect fluid'' consistent with quantum field theory?  }
\author{Thomas D. Cohen }
\affiliation{Department of Physics \\
University of Maryland\\College Park, MD 20742}

\begin{abstract}
It was recently conjectured that the ratio of the shear viscosity
to entropy density, $ \eta/ s$, for any fluid always exceeds
$\hbar/(4 \pi k_B )$. This conjecture was motivated by quantum
field theoretic results obtained via the AdS/CFT correspondence
and from empirical data with real fluids. A theoretical
counterexample to this bound can be constructed from a
nonrelativistic gas by increasing the number of species in the
fluid while keeping the dynamics essentially independent of the
species type. The question of whether the underlying structure of
relativistic quantum field theory generically inhibits the
realization of such a system and thereby preserves the possibility
of a universal bound is considered here. Using rather conservative
assumptions, it is shown here that a metastable gas of heavy
mesons in a particular controlled regime of QCD provides a
realization of the counterexample and is consistent with a
well-defined underlying relativistic quantum field theory. Thus,
quantum field theory appears to impose no lower bound on $\eta/s$,
at least for metastable fluids.
\end{abstract}

\maketitle

A universal bound for  the ratio of the shear viscosity, $\eta$,
to entropy density, $s$, in {\it any} fluid has recently been
conjectured by Kovtun, Son and Starinets (KSS):
\begin{equation}
\frac{\eta}{s}\ge\frac{\hbar }{4 \pi k_B} \label{strong}
\end{equation}
where the $k_B$ and $\hbar$ are Boltzmann's constant and  Planck's
constant, respectively \cite{KSS1}.  This is the strong form of
the KSS conjecture.   Since the limit of zero viscosity is the
``perfect'' fluid, the bound from the strong KSS conjecture
implies that ``the most perfect'' fluid is one with $\eta = {\hbar
\, s}/{(4 \pi k_B)}\, $.  For simplicity, in the remainder of this
letter units will be chosen using standard theory conventions with
$\hbar$ and $k_B$ set to unity. The KSS conjecture in its strong
form represents an extremely important advance in our
understanding of many-body physics, if correct. Indeed, it has
been invoked in discussing physics as diverse as ultracold gases
of trapped atoms\cite{AMO} to the quark-gluon plasma\cite{QGP}.

A fluid is a substance that continually deforms ({\it i.e.} flows)
under an applied shear stress regardless of the magnitude of the
applied stress. Usually, this definition is taken to hold
regardless of whether the state of the system is stable or
metastable (in the sense of being very long-lived on all relevant
time scales). Consider a liquid high explosive---which obviously
is not absolutely stable. Such a substance is usually regarded as
unambiguously being a fluid. In principle, $\eta$ and $s$ can be
measured for a liquid high explosive with extremely high accuracy
despite its lack of absolute stability (although great care is
advised in making the measurements). There may be a fundamental
uncertainty in values of $\eta$ and $s$ arising from a natural
time scale for the lifetime of the metastable state, $\tau_{\rm
met}$. However, one expects that any fundamental uncertainty to be
of relative order, ${\tau_{\rm fl}}/{\tau_{\rm met}}$, where
$\tau_{\rm fl}$ is the natural time scale of the fluid dynamics
({\it e.g.} the collision time for a dilute system). Since, this
ratio is many orders of magnitude less than unity for typical
metastable fluids, the entropy density and viscosity are
essentially well defined. It is natural to suppose that if the KSS
bound is a generic feature of fluids that it will also hold for
metastable fluids (up to the very small intrinsic uncertainty in
the value of $\eta$ and $s$ due to metastability), although it is
logically possible that the bound holds only for stable fluids.

It is useful at the outset to review the arguments which make the
conjecture plausible. A principal motivation is based on the
transport properties on gauge theories with gravity duals
computed\cite{transport} using the famed AdS/CFT
correspondence\cite{AdSCFT}. A truly remarkable result has
emerged\cite{KSS1}.  For systems at nonzero temperature, at large
$N_c$ and infinitely strong `t Hooft coupling $g^2 N_c$, all known
theories with a gravity dual have the same  ratio of the shear
viscosity ($\eta$) to the entropy density ($s$):
$\frac{\eta}{s}=\frac{1}{4 \pi}$ . This suggests a general feature
of classical gravity which should hold for all large $N_c$
theories with gravity duals; in refs.~\cite{uni,KSS2} it was shown
to hold in a very wide class of theories.  It is natural to assume
that $\eta/s$  increases as one departs from the limit of infinite
strong `t Hooft coupling since weaker couplings should increase
$\eta$. An explicit calculation confirmed this behavior for a
specific model\cite{ho}. Thus, a class of large $N_c$ gauge
theories with gravity duals appears to have a bound for $\eta/s$.
The KSS conjecture is that the bound could be much more general.
Moreover, the conjectured bound does not involve $c$ (when units
are restored) and it is not unnatural that the bound, if true,
applies equally to nonrelativistic systems\cite{KSS1}.  A
heuristic argument for the bound in nonrelativistic systems can be
found in ref.~\cite{GSZ}.

The bound is respected by all real fluids
considered\cite{KSS1,KSS2}. Typically, for a fluid at fixed
pressure and low temperatures, $\eta/s$ decreases with increasing
temperature until a minimum is reached and then increases.  The
minima all appear to be well above the putative bound, providing
an empirical basis for the conjecture.


However, in a second publication\cite{KSS2} KSS briefly note a
serious threat to this universal conjecture based on rather
elementary theoretical considerations indicating a class of
counterexample. The issue raised there is critical to the
discussion and an expanded variant of this argument is given
below.

Consider a nonrelativistic quantum many-body system in a regime
for which the computation of $\eta/s$ is analytically tractable.
The system consists of a gas comprised of a number ($N_s$) of
distinct but equivalent species of spin-zero bosons of degenerate
mass. The dynamics is specified by a Hamiltonian containing
kinetic energy terms plus two-body interactions which are
independent of the species type: for all species $a$ and $b$,
$V_{a b} (r) = \, V(r/R)$ where $R$ is the range of the potential.
The gas is in local thermal equilibrium and has an equal density
of each species: $n_a = n/N_s$ for any species $a$.  The system is
in the regime
\begin{equation}
R^{-2} \, , a^{-2} \gg m T \gg n^{2/3} \label{regime}
\end{equation}
where $a$ is the scattering length. This simultaneous low
temperature, low density regime can be enforced by a common
scaling behavior with a parameter $\xi$:
\begin{equation}
n = \frac{n_0}{\xi^4} \; \; \; T=\frac{T_0}{\xi^2} \; .
 \label{scale1}
\end{equation}
Sufficiently large $\xi$ ensures the regime of Eq.~(\ref{regime}).
At large $\xi$, $n$ is low enough for the entropy to be given by
the classical ideal gas (with small corrections). Moreover, the
many-body dynamics is dominated by binary collisions and hence is
in the regime of validity of the Boltzmann equation\cite{KT}. The
low temperature aspect implies that the two-body scattering
amplitude of a typical pair is predominantly s-wave and (to good
approximation) equal to the scattering at zero momentum. Thus the
scattering term in the Boltzmann equation is based on isotropic
scattering independent of energy.  This is formally equivalent to
the Boltzmann equation for {\it classical} hard sphere scattering;
a case for which the shear viscosity is readily computed: $\eta=
C_{\rm hs}\sqrt{m T}/d^2$ where $d$ is the diameter and $C_{\rm
hs} \approx .179$ is a coefficient calculable
numerically\cite{KT}. The ratio $\eta/s$ is universal in this
regime and is given by:
\begin{equation}
\frac{\eta}{s} =   \, \frac{C_{\rm hs} \,  \xi^3 \, \sqrt{  m \,
T_0} } {
 a^2 \, n_0 \left (  \log \left(\frac{ (m \, T_0)^{3/2}}{n_0}
\right )+ \frac{5}{2} +  \log (\xi) + \log (N_s) \right ) }\; .
\label{result1}
\end{equation}

Corrections to Eq.~(\ref{result1}) are suppressed by various
powers of $1/\xi$ and should be irrelevant at sufficiently large
$\xi$. The validity of Eq.~(\ref{result1}) at large $\xi$ should
not depend on the number of species being small. Accordingly it is
legitimate to consider an exponentially large number of species:
take $N_s$ to be
\begin{equation}
N_s = \exp ( \xi^4) \;  \label{scale2}
\end{equation}
(rounded to the nearest whole number). As the temperature and
density of the system is decreased, the number of species
simultaneously increases greatly. In the combined regime of
Eqs.~(\ref{scale1}) and (\ref{scale2}), $\eta/s$ is given as:
\begin{equation}
\frac{\eta}{s} = \,\frac{1}{\xi} \,  \frac{C_{\rm hs} \sqrt{ m \,
T_0} } {  a^2 \, n_0 } \, \label{result2}
\end{equation}
up to power law corrections. Clearly $\eta/s$ can be made
arbitrarily small by increasing $\xi$ and, in particular, can be
made smaller than $1/(4 \pi)$. At large $\xi$, the mixing entropy
associated with the many species overwhelms other effects yielding
a small value of $\eta/s$.

For purely repulsive potentials the argument appears to hold
generally. As a practical matter it would be very difficult to
realize this example in realistic circumstances---the number of
equivalent species must be the exponent of a large number.
However, it does show as a matter of principle that nothing in
ordinary many-body quantum mechanics requires that the conjectured
strong KSS bound be respected.

Subtleties arise if the potential has attractive regions.  The
second derivative of the free energy density is given by, $
\frac{d^2 f}{d n^2} = \xi^2 \frac{T_0}{n_0} - 4 \pi |a|/m$  (up to
power law corrections in $\xi^{-1}$). It is clearly positive for
large $\xi$ indicating that the fluid is {\it locally} stable
against density fluctuations. However, it may be possible for the
system to globally lower its energy by either forming macroscopic
regions of higher density or by forming two- or higher-body bound
states. Regardless of whether the system is absolutely stable, the
time scale for the decay of the system is very long at large
$\xi$. The fastest possible mechanism for the system to
qualitatively alter its state is if two-body bound states exist.
To conserve energy and momentum, two particles can only bind if
three or more particles interact.  This implies that in this
situation ${\tau_{\rm fl}}/{\tau_{\rm met}} \sim n R_0^3 \sim
\xi^{-4}$, and the system becomes a well-defined fluid at large
$\xi$.  For other mechanisms, the lifetime of the metastable
system is parametrically larger. Thus, regardless of whether the
interaction is attractive or repulsive a fluid violating the KSS
bound can be constructed (although for an attractive interaction
it may only be metastable.)

It is useful to introduce another variant of this example which
will be of use later.  Suppose that in addition to $N_s$, $n$ and
$T$, the mass and the strength of the interaction potential also
have a nontrivial scaling with $\xi$:
\begin{eqnarray}
m & = & m_0 \exp(\xi^4)  \; \; \; \;   R=R_0 \; \; \;  \; V_{a
b}(r) =
{V_o(r/R_0)}\exp(-\xi^{4}) \nonumber \\
T& =& \frac{T_0 \exp (-\xi^{4)}}{\xi^2} \; \;  \; \; \; \; \;
n=\frac{n_0}{\xi^4} \; \; \; \; \; \; \; N_s= \exp (\xi^4) \; .
\label{scale3}
\end{eqnarray}
In this scaling, the particles become heavy at the same rate as
the number of species grows, while their interaction potentials
become weak at the same rate. The two-body scattering depends only
on the product $m V$ and is unaltered by this scaling; thus, the
cross section remains independent of $\xi$.  Note, that $T$ in
Eq.~(\ref{scale3}) drops off exponentially faster than the
analogous scaling in Eq.~(\ref{scale1}). However, the relevant
combination is the thermal momentum, $\sqrt{m T}$, which scales as
$\xi^{-1}$ in both scaling regimes. Thus, for this scaling regime
$\eta /s$ is still given by Eq.~(\ref{result2}) with $m$ replaced
by $m_0$; corrections are still down by powers of $\xi^{-1}$ and
the KSS bound is still violated at large $\xi$.

One way to accommodate counterexamples  of this sort is to discard
the strong form of the conjecture in favor of a much weaker
version---any relativistic quantum field theoretic system at zero
chemical potential has the ratio of $\eta/s \ge (4 \pi)^{-1}$---as
proposed by KSS in ref. \cite{KSS2}. The weaker form of the
conjecture remains of great theoretical importance, if true.
However, its practical implications are severely limited; apart
from relativistic heavy ion collisions where the chemical
potential may be low enough to neglect, the vast majority of fluid
problems of physical interest are outside of its domain. Moreover,
the weak form of the conjecture seems less plausible if the the
strong form is discarded. A principal justification for the
conjecture is empirical data on ordinary fluids; this provides no
support for the weak version. In addition, it has recently been
shown\cite{mu} that systems with gravity duals at infinite `t
Hooft coupling and large $N_c$ with non-zero chemical potentials
also have $\eta/s = 1/(4 \pi)$ just as with zero chemical
potential. To the extent that the zero chemical potential case
motivates the weak form of the conjecture, one might expect a
similar level of justification to hold at non-zero chemical
potential.

However, the very fact that the counterexamples are so impractical
raises an obvious question: could there be some deep physical
reason of principle preventing them from being realized?  This
could preserve the strong form of the conjecture.  Since this
physics must be  beyond the structure of nonrelativistic quantum
mechanics, it is natural to ask if the structure of the underlying
relativistic quantum field theory could do this. The issue is
whether it is possible for nonrelativistic systems inconsistent
with the strong KSS conjecture to arise from any consistent
underlying quantum field theory; {\it i.e.}, whether there exists
a consistent ultraviolet completion for the nonrelativistic
system\cite{DTS}.

{\it A priori} it is not implausible that the structure of quantum
field theory  might rule out these counterexamples and preserve
the strong form of the conjecture. Consider a simple example: a
low temperature, low density pion gas in a generalization of QCD
with $N_f$ degenerate flavors---half of which we denote to
be``up-like'' and half ``down-like''. The net number density for
up-like quarks and down-like anti-quarks can be set to $n$ and the
temperature can be chosen to be well below $m_\pi$, yielding a
nonrelativistic pion gas with $N_f^2/4$ species of pion present.
However, it is not possible to take $N_F$ large (and thus evade
the KSS bound) while keeping the theory well defined. The one-loop
$\beta$ function for QCD is $ \beta(g) = - \frac{g^3}{16 \pi^2}
\left(\frac{11 N_c}{3} - \frac{2 N_f}{3} \right ) $;  asymptotic
freedom is lost at large $N_f$ and the theory presumably becomes
ill defined in the ultraviolet\cite{pion}.  One might evade this
by setting $N_c$ large simultaneously with $N_f$ with the ratio
$N_f/N_c$ held fixed. However, $\pi-\pi$ scattering cross section
scales as $N_c^{-2}$ implying that $\eta \sim N_c^2$.  Thus,
either the theory is ill defined in the ultraviolet or $\eta/s$
grows with increasing species number. If all consistent field
theories behave in an analogous way, then the strong KSS
conjecture would remain viable. However, this behavior is not
generic.

Consider a nonrelativistic gas of heavy mesons  ({\it i.e.},
mesons with the quantum numbers of a heavy quark and a light
anti-quark) in a very special regime of generalized QCD with weak
and electromagnetic interactions absent. The regime involves a
theory with $N_c$ colors, $N_h$ degenerate flavors of heavy quark
(of mass $m_h$), and one flavor of light quark (of mass $m_l$).
The total density of light anti-quarks to be $n$ and the density
of each heavy flavor is fixed at $n/N_h$ (for a total heavy meson
density of $n$). The parameters of the theory are taken to scale
according to:
\begin{eqnarray}
&{}& N_c   =  e^{\xi^4} \;  \;   N_h =  e^{\xi^4} \;  \; \; \; m_h
= {m_h}_0 \, e^{\xi^4} \; \; \;  \;   m_l \sim  {m_l}_0 \nonumber
\\
&{}& \Lambda_{\rm QCD}  =   {\Lambda_{\rm QCD}}_0 \; \; \; \;  \;
\; n = n_0 \xi^{-4} \; \; \;  \; \;  T = T_0
\frac{e^{-\xi^4}}{\xi^2} \,
 . \label{scale4}
\end{eqnarray}
The scaling relations in Eqs.~(\ref{scale4}) are designed to
create a nonrelativistic gas of heavy mesons in a regime
equivalent to that of Eq.~(\ref{scale3}).   The problem of
diminishing cross section in the large $N_c$, large $N_f$ pion gas
is evaded here by having the meson mass grow to compensate for the
decreased interaction strength yielding a constant cross section.

There are several aspects of this system which require comment.
First, note that the scaling rules give  $N_f=N_h+1=N_c+1$.
Asymptotic freedom is maintained and the theory remains well
defined at large $\xi$ (assuming, of course, that QCD itself is,
in fact, well defined).

The next key point is that the system  behaves like a
nonrelativistic gas whose constituents are the lightest
pseudoscalar mesons containing a heavy quark.  The pseudoscalar
containing a quark of flavor $a$ is denoted as $H_a$. Recall that
for heavy quarks, the vector meson $H_a^{\star}$ is nearly
degenerate\cite{IW} with the pseudoscalar $H_a$: $(M_{H}^\star-
M_{H}) \sim \frac{\Lambda_{\rm QCD}^2}{m_{H}} \sim e^{-\xi^4}
\,\frac{\Lambda_{QCD}^2}{{m_h}_0}$. Combining this with the
scaling of the temperature, one sees that $\frac{M_H^{\star}-
M_H}{T} \sim \xi^2 \, \frac{\Lambda_{\rm QCD^2}}{{m_h}_0}$ so that
at sufficiently large $\xi$, excitations of the $H^\star$ are
thermally suppressed; the probability of a heavy meson being in
the $H^\star$ state is exponentially suppressed by a factor of
$e^{-\xi^2}$. Other excited states are suppressed by {\it much}
larger factors ({\it i.e.}, by exponentials of exponentials of
$\xi$). Similarly, the density of light mesons in the system is
thermally suppressed by a very large factor at large $\xi$. Thus,
both the viscosity and the entropy of the system are dominated by
the $H$ mesons. Moreover, the $H$ mesons are clearly extremely
nonrelativistic: the characteristic thermal velocity $v_T =
\sqrt{T/M_H}$ scales as $e^{-\xi^4}/\xi^2$.

Only the dynamics of $H$ mesons are relevant at large $\xi$; all
other degrees of freedom can be integrated out yielding an
effective theory.  The appropriate effective theory is
nonrelativistic quantum mechanics with an effective potential
between the $H$ mesons; the particle number is essentially fixed
and the motion is nonrelativistic. From standard large $N_c$
analysis it is easy to see that the strength of the effective
potential scales like $1/N_c$; the range of the effective
potential is fixed from the {\it light} degrees of freedom (light
quark and gluons) since virtual heavy quarks may be integrated out
in for large $\xi$.
Thus, in the regime of Eq.~(\ref{scale4}), the system behaves like
a nonrelativistic gas of pseudoscalar heavy mesons; their
properties scale analogously to those in Eqs.~(\ref{scale3}). From
the logic leading to Eq.~(\ref{result2}) one expects that
$\frac{\eta}{s} = \,\frac{1}{\xi} \,  \frac{C_{\rm hs} \sqrt{
{m_h}_0 \, T_0} } { a^2 \, n_0 }$ where $a$ is the meson-meson
scattering length (corrections are power law suppressed in
$1/\xi$).  The KSS bound is violated at large $\xi$ since $\eta/s$
becomes arbitrary small.

Given this counterexample, it appears that the strong form of the
KSS conjecture cannot be supported:  a fluid based on an
underlying consistent relativistic field theory can have an
arbitrarily small value of $\eta/s$.  Of course, the preceding
argument does not rule out the strong form of the KSS conjecture
in any rigorous mathematical sense.  The calculation of $\eta/s$
in this regime was based on numerous approximations and is not an
exact solution of QCD (which, in turn has never been shown to
exist in a rigorous mathematical way). However, these
approximations appear to be well motivated physically and the
result should be exact in the limit of large $\xi$; accordingly,
it seems highly unlikely that there exists a fundamental bound on
the ``the most perfect'' fluid which is consistent with an
underlying quantum field theory.

There is an important caveat to this conclusion. In the heavy
meson system, there are attractive interactions between, and the
fluid can be shown to be metastable rather than stable (with
${\tau_{\rm fl}}/{\tau_{\rm met}}$ going to infinity at large
$\xi$). The logical possibility exists that the strong KSS
conjecture holds but only for stable fluids. {\it A priori} it is
unclear how plausible it is for the KSS bound to hold for stable
fluids, but not to hold (even approximately) for metastable fluids
no matter how long-lived. Moreover, if the strong KSS bound only
holds for absolutely stable systems, it cannot be applied directly
to ordinary fluids such as water which are {\it not} stable under
the standard model.  In principle the free energy of water can be
lowered via nuclear reactions ({\it e.g.}, two hydrogen nuclei
combining to from a deuteron). One cannot argue that the standard
model dynamics are irrelevant (on the grounds that the dynamics
are essentially purely electromagnetic in nature); the logic
supporting the strong KSS conjecture depended on the existence of
an underlying quantum field theory. Thus, the direct
application\footnote[1]{D.T. Son has pointed out in a private
communication that it may be possible to apply the bound {\it
indirectly} for ordinary fluids. The idea is to construct a
quantum field theory with electrons and fundamental fields for
nuclei and with electromagnetic interactions which look like those
in the standard model and therefore yields essentially the same
fluid properties. However, it is unclear whether such a theory can
be found which is both well defined and yields absolutely stable
fluids.} of the strong KSS bound to ordinary fluids is legitimate
only if the bound applies to metastable fluids; the previous
counterexample implies that it does not.

The author has greatly benefited from numerous insightful comments
of D.T. Son. The support of the U.S. Department of Energy is
gratefully acknowledged.

\end{document}